\documentclass[conference]{IEEEtran}
\usepackage{url}
\usepackage{amsthm}
\usepackage{amsmath}
\usepackage{comment}
\usepackage[caption=false]{subfig}
\usepackage{cite}

\newtheorem{finding}{Finding} 
\newtheorem{researchquestion}{Motivating question} 
\usepackage{comment}
\usepackage[english]{babel}
\usepackage[utf8]{inputenc}
\usepackage{tabularx}
\usepackage{graphicx}

\begin{document}

\newcommand{\timetolive}{T}
\newtheorem{theorem}{Theorem}
\newtheorem{lemma}{Lemma}
\newcommand{\costmon}{C_{\textrm{mon}}}
\newcommand{\nmon}{N_{\textrm{mon}}}
\newcommand{\nmiss}{N_{\textrm{miss}}}
\newcommand{\nmoni}{N_{\textrm{mon}}^{(i)}}
\newcommand{\nmissi}{N_{\textrm{miss}}^{(i)}}
\newcommand{\costmiss}{C_{\textrm{miss}}}

\title{Learning When to Say Goodbye: What Should be the Shelf Life of an Indicator of Compromise?}

\author{\IEEEauthorblockN{Breno~Tostes\IEEEauthorrefmark{1},        Leonardo~Ventura\IEEEauthorrefmark{1},   Enrico~Lovat\IEEEauthorrefmark{3},  Matheus~Martins\IEEEauthorrefmark{3},  Daniel~Menasché\IEEEauthorrefmark{1}} 
\IEEEauthorblockA{{\IEEEauthorrefmark{1}Federal University of Rio de Janeiro (UFRJ), }
 {\IEEEauthorrefmark{3}Siemens Corporation } }}

\maketitle

\begin{abstract}
Indicators of Compromise (IOCs), such as IP addresses, file hashes, and domain names associated with known malware or attacks,  
are cornerstones of cybersecurity, 
 serving to identify   malicious activity on a  network. 
In this work, 
we leverage real data to compare different parameterizations of IOC aging models. Our dataset comprises traffic at a real environment for more than 1 year.   Among our trace-driven findings, we determine thresholds for the ratio between miss  over monitoring costs such that the system benefits from storing IOCs for a finite time-to-live (TTL) before eviction.   To the best of our knowledge, this is the first real world evaluation of thresholds related to IOC aging, paving the way towards realistic 
   IOC decaying models.
\end{abstract}

\begin{IEEEkeywords}
Threat intelligence; data science; modeling and analysis.
\end{IEEEkeywords}
\section{Introduction}\label{sec:introduction} 

Indicators of Compromise (IOCs), such as IP addresses of compromised hosts, hashes of malware and bodies of emails of phishing campaigns,  are the foundation of cyber threat intelligence (TI). They serve as signatures of risk, being  employed in monitoring systems to generate alerts if a match is found  between known IOCs and  data collected at a given environment.   In essence, the larger the IOC base, the greater is the coverage against previously observed cyber-attacks.   However, such coverage  is associated  with its own costs~\cite{bouwman2020different}.

The security of the target environment could be at risk if the  number of monitored IOCs is limited, as it may lead to the omission of crucial indicators.
%
%
 %
On the other hand, maintaining too many IOCs is prohibitive due to intrinsic  costs of  investigating a large catalog of potential incidents~\cite{liao2016acing}.  Understanding the dynamics of IOC creation and sightings is crucial for addressing the challenges of maintaining and using IOCs,  and for developing effective strategies for monitoring and responding to cyber threats.

\begin{figure}
    \centering
    \includegraphics[width=0.8\columnwidth]{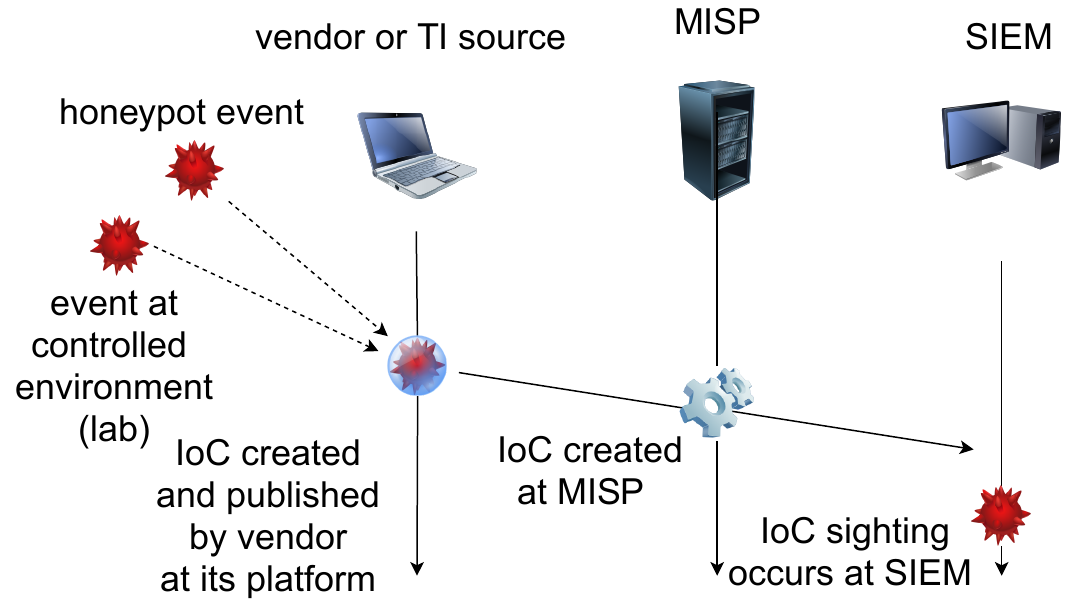}
    \caption{Ecosystem of vendors, MISP instances and SIEMs sharing   IOCs and   their sightings.} 
    \label{fig:IOC_dynamics}
\end{figure}

Figure~\ref{fig:IOC_dynamics} shows the typical dynamics of IOC creation and sightings.  An IOC is typically discovered at a vendor or TI source, e.g., in a controlled lab environment or through a honeypot.  Then, the IOC is  created and published by that  vendor, and is also   propagated to sharing platforms, such as the  Malware Information Sharing Platform (MISP).  MISP is a  distributed system comprising multiple instances, run by different organizations and communities, and   allowing them to benefit from the collective knowledge. 
Finally, Security Information and Event Management (SIEM) systems monitor the IOCs, and eventually report sightings for the monitored IOCs at the Security Operation Centers (SOCs).

Over time, IOCs lose relevance and their monitoring  leads to
costs  
due to outdated information.  
Indeed, numerous false positives challenges  cognitive limitations of  SOC employees, and builds   monetary costs that vary depending on    monitoring  price models.  Azure Sentinel Threat Intelligence, for instance, offers two pricing alternatives: Capacity Reservations
 and Pay-As-You-Go. In the latter,  the current cost is of 	\$2.46 per GB-ingested.  If the presence of IOCs is used to pre-filter   data to be fed to Azure,  the larger the number of IOCs being monitored, the larger the incurred costs~\cite{sentinel}.

To cope with the aging of IOCs, threat intelligence platforms, such as   MISP, have introduced models to determine when an IOC should no longer be monitored. Those models, referred to as aging  models (or decaying models), however, have a number of parameters whose assignment poses its own challenges,  thus motivating  explainable models with simpler  and   fewer  parameters~\cite{iklody2018decaying}.
In particular, we focus on a model with  two main parameters, namely the missing and monitoring costs, related to the impact
of missing a sighting and to the attention span
required to handle alarms, respectively. 
Our study aims to answer two key questions: How long should a certain indicator be monitored for, and what is the optimal aging model parameterization for a given environment? 

\begin{figure}[t]
    \centering
    \includegraphics[width=\columnwidth]{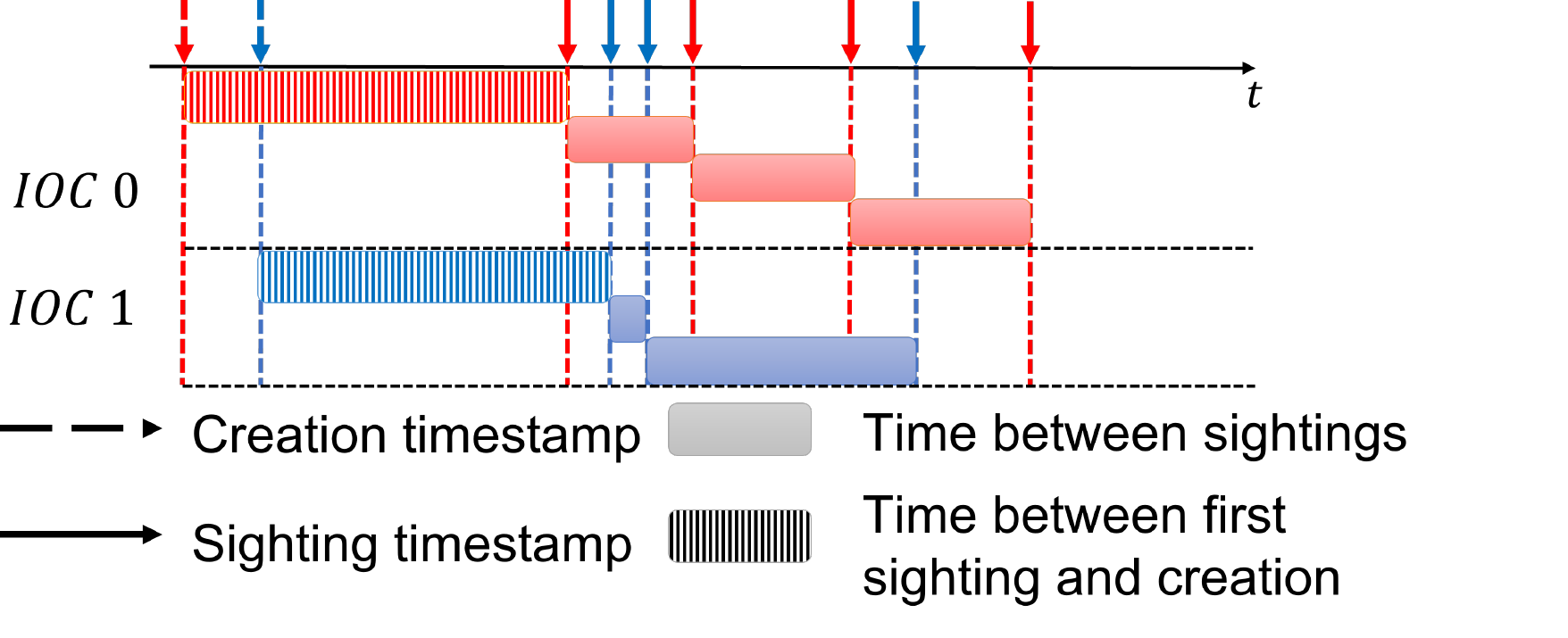}
    \caption{Dynamics of IOC   creations   and  sightings.}
    \label{fig:IOC_dynamics2}
\end{figure}

To answer the above questions, we  leverage real data to compare different parameterizations of IOC aging models.  
Our dataset comprises  traffic from a real-world enterprise environment spanning over a year. 
%
%
It contains for each IOC the instants at which sightings occurred  (see Figure~\ref{fig:IOC_dynamics2}).  Those sightings are used to parametrize aging models.


In the MISP terminology, an event refers to a collection of related IOCs.  A given IOC may be part of multiple events, but each sighting corresponds to a single event. 
  In summary,
    our dataset contains, for each sighting towards each IOC,  its 1)  timestamp, 2)   IOC anonymized identifier,
3) 
            IOC type, e.g., domain, IP source, IP destination, email source, email subject, md5, sha1, sha256, filename, hostname or URL,  
    4) IOC creation date, 
      and   5)  event identifier. 
         Note that data is anonymized, so the event is characterized by a  non-informative identifier, that serves solely to determine how sightings are related to each other through their corresponding events.

         %
         %

Using our dataset, we characterize how the number of sightings towards IOCs varies over time, as a function of IOC types, and assess the impact of  aging model parameters on coverage and corresponding monitoring costs~\cite{romanosky2016examining}.
%
  %
  %
  %
%
%
%
The 
\emph{hit ratio}  corresponds to  the fraction of IOC sightings that occur while the corresponding IOC is being monitored.  A sighting to an unmonitored IOC  is said to be uncovered, contributing towards its \emph{miss ratio}.   Correspondingly, the \emph{monitoring cost} at any given point in time is proportional to the number of IOCs that are being monitored at that time.


\textbf{Contributions. }  Our key contributions are twofold.




\textbf{Formulation of the TTL optimization problem. } 
 To each IOC we associate a corresponding TTL.  The TTL  is initialized at a constant value, and is decremented at every time unit. When TTL reaches zero, the corresponding IOC monitoring is discontinued.   Such a TTL decaying model has a number of different flavors~\cite{dehghan2019utility,mokaddem2019taxonomy,ermerins2020scoring}. 
 Under TTL with reset, the TTL is reset to its initial value  whenever a sighting occurs.   Under TTL without reset, in contrast, sightings do not impact the TTL dynamics.  In any case, note that the TTL dynamics are decoupled across multiple IOCs. 
 %
 %
We let $\timetolive$  denote the initial TTL value, and show how $\timetolive$ impacts monitoring costs and miss ratio, under TTL with and without reset. 

\textbf{Trace driven findings. }
Among our trace-driven findings, we discover that 
if the cost of  missing a sighting  is below 2,152 times the daily cost of monitoring an IOC, it is not worth incurring the monitoring costs for any IOC. Conversely, if the cost of a miss is beyond $10.5$ million dollars,  all IOCs should be constantly monitored, assuming a  unitary daily dollar cost for  monitoring an IOC.  For values inbetween  those two thresholds, 
%
 the system benefits from storing IOCs for a finite time-to-live (TTL), which can be set according to the IOC category. For instance, if the TTL is set to 248 days then the sum of miss costs and monitoring costs is minimized when the cost of  missing a sighting  equals 10,000 times the daily cost of monitoring an IOC.  To the best of our knowledge, this is the first real world evaluation of IOC aging thresholds.
 
 \textbf{Outline.} The remainder of this paper is organized as follows.  In Section~\ref{sec:trace} we introduce our trace. Then, Sections~\ref{sec:deterministicandoutliers} and~\ref{sec:categories} report our trace-driven findings.  Section~\ref{sec:utilities} introduces a utility optimization approach to set  TTLs  and reports our model-driven findings. Finally, Section~\ref{sec:conclusion} concludes. 
 

\begin{figure}[t]
    \centering
\includegraphics[width=0.7\columnwidth]{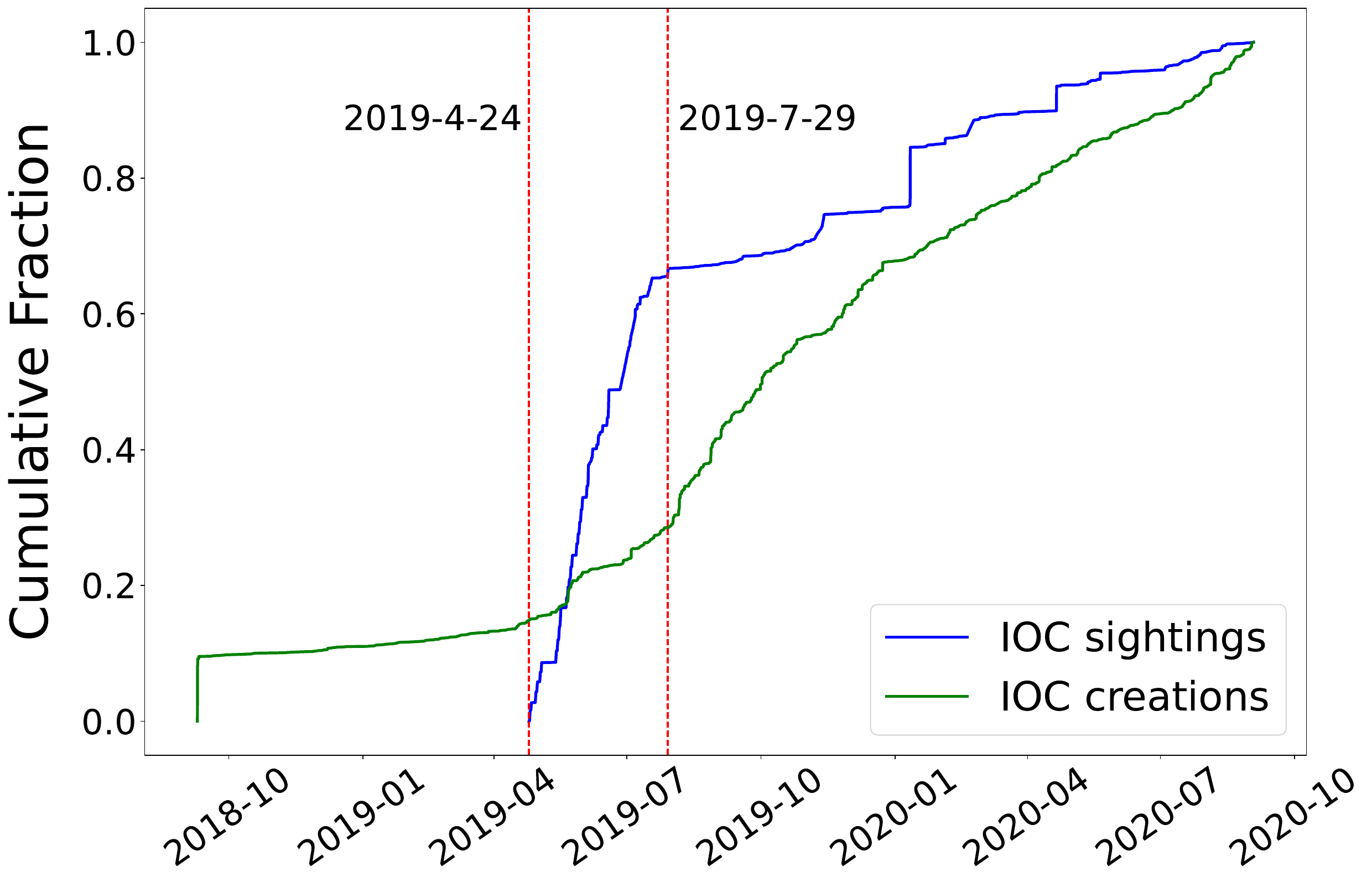}
    \caption{Cumulative fraction of IOCs created and   sightings}
    \label{fig:cdf_IOC}
\end{figure}

\section{Dataset Description} \label{sec:trace}

Our trace was collected from a  SOC of a large scale company, comprising more than 300,000 employees and gathering data from more than 12 countries, and contains 5,789 IOCs with at least one sighting.  The first and last IOC creation dates occurred at 9/9/2018 and 9/2/2020.\footnote{For the considered company,  sensitive information  becomes unclassified after 2 years, as far as it is anonymized.}  The last sighting  occurred at the same date at which  the last IOC was created. 
Figure~\ref{fig:cdf_IOC} shows the cumulative fraction of   IOCs created during the interval of our trace,   and the cumulative fraction of  sightings    issued across that interval.  Among all sightings,  66\%  occurred  during the  first three months of our monitoring campaign, between 4/24/2019 and 7/29/2019 (see Table~\ref{tab:general}).

It is worth noting that some IOCs may have sightings before their corresponding creation dates. This is because once an IOC is created, one may, in retrospect,  
detect  occurrences of the IOC  at  system  logs. 
Among the   IOCs with sightings in our trace, 9\% (530/5,789)  contain sightings one day before their creation dates (see Figure~\ref{fig:pie_chart}).  In Appendix~\ref{sec:appworkload} we provide a more detailed statistical characterization of time between sightings and creation dates,  leveraging survival analysis for this purpose. 




\begin{researchquestion}How  are IOCs related to each other?
\end{researchquestion}

\begin{figure}[!t]
    \centering
    \subfloat[Number of sightings per IOC type]{
        \includegraphics[width=0.5\columnwidth]{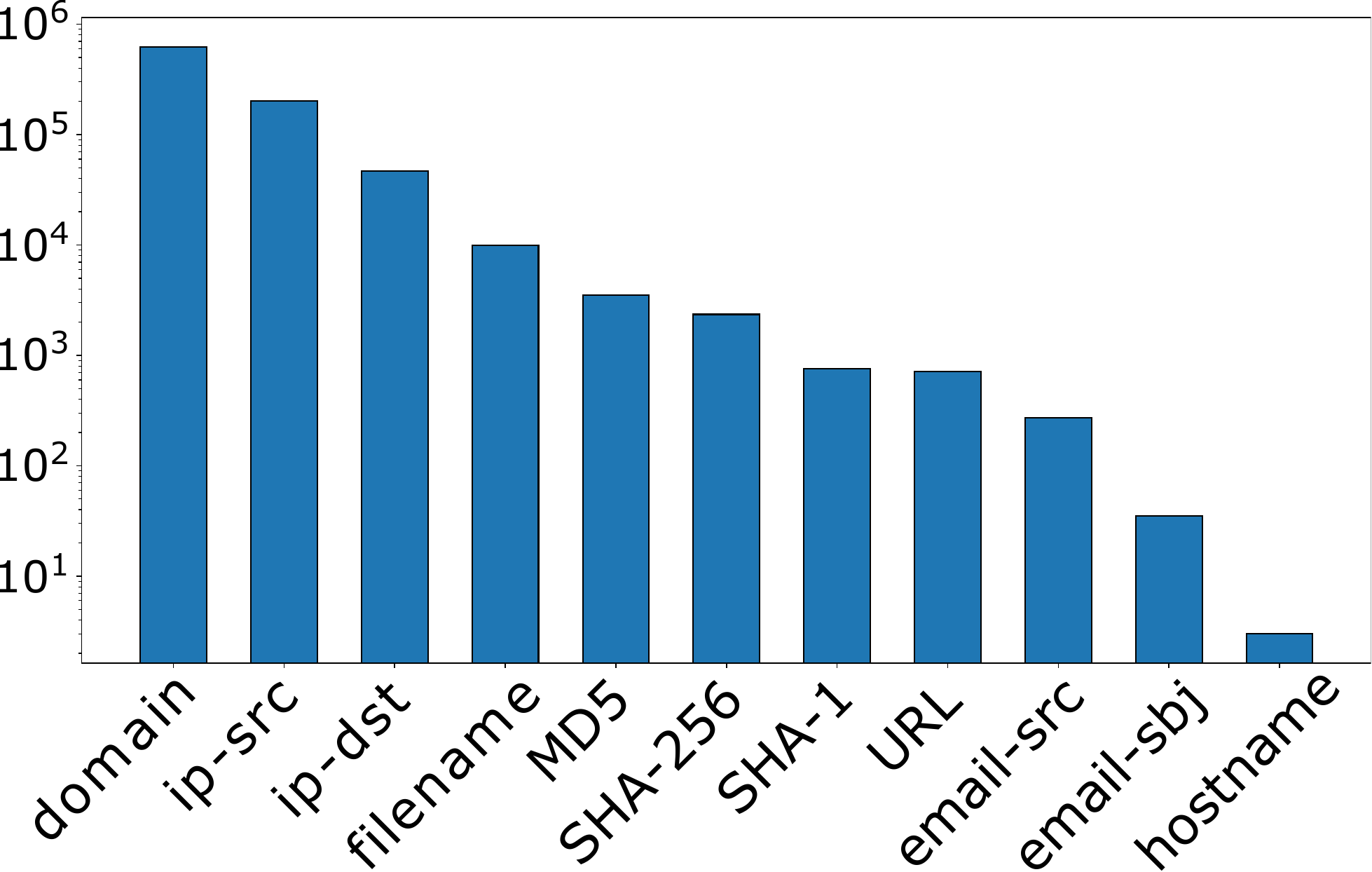}
        \label{fig:sightings_per_type}
    }
    \subfloat[Number  of IOCs per IOC type]{
        \includegraphics[width=0.5\columnwidth]{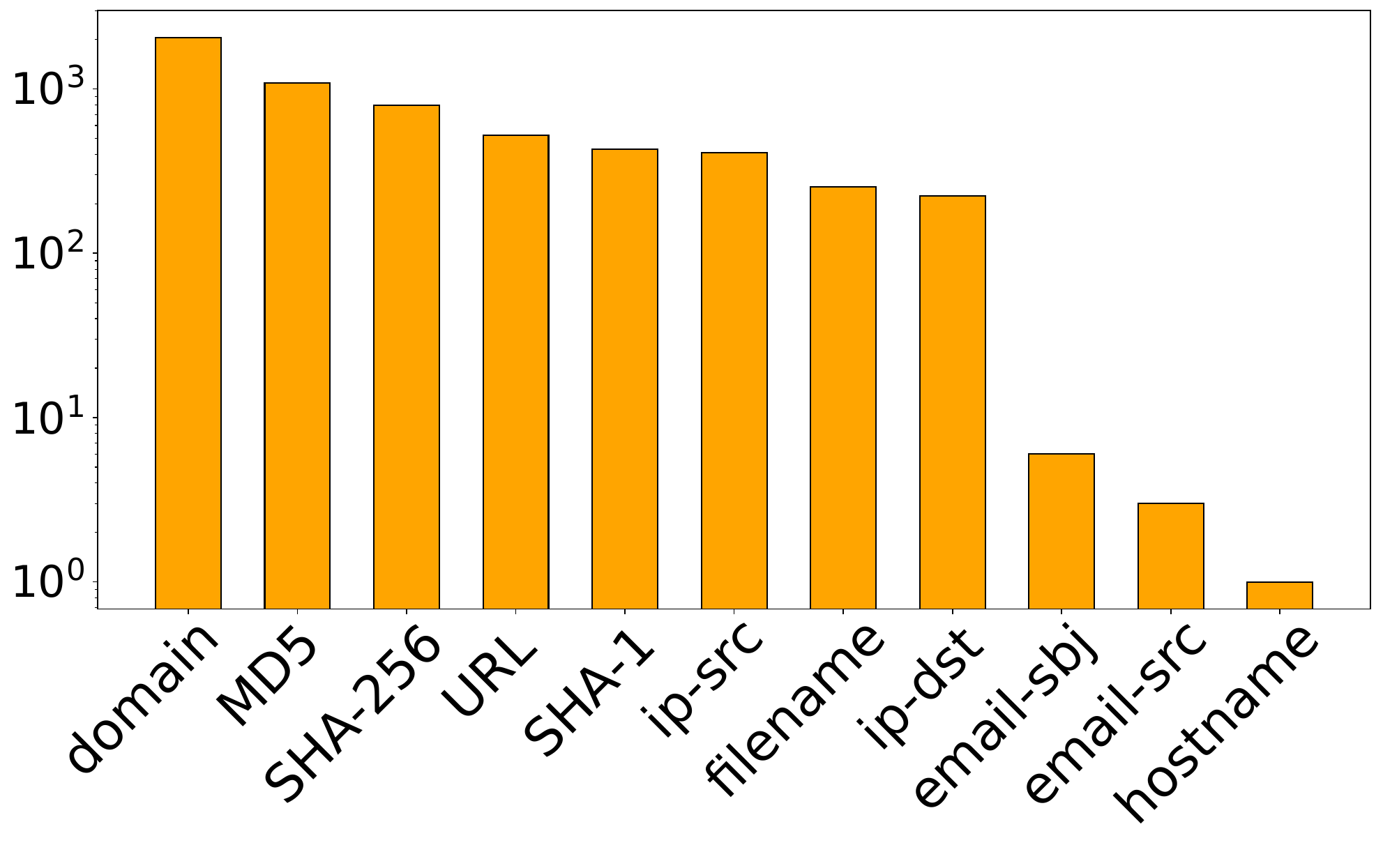}
\label{fig:iocs_per_type}
 }
    \caption{Number of sightings and     IOCs per   type. Domains correspond to the vast majority of sightings and IOCs.   IPs also yield a significant fraction of sightings, while hashes have   the second-largest number of IOCs. }
    \label{fig:type_statistics}
\end{figure}

In our trace, each IOC belongs to one of the eleven types shown in Figure~\ref{fig:type_statistics}.  Most of the observed indicators and IOCs are associated with domains. IPs also contribute to a considerable proportion of sightings, whereas hashes have the second-highest number of IOCs.

\begin{figure}
    \centering
    \includegraphics[width=\columnwidth]{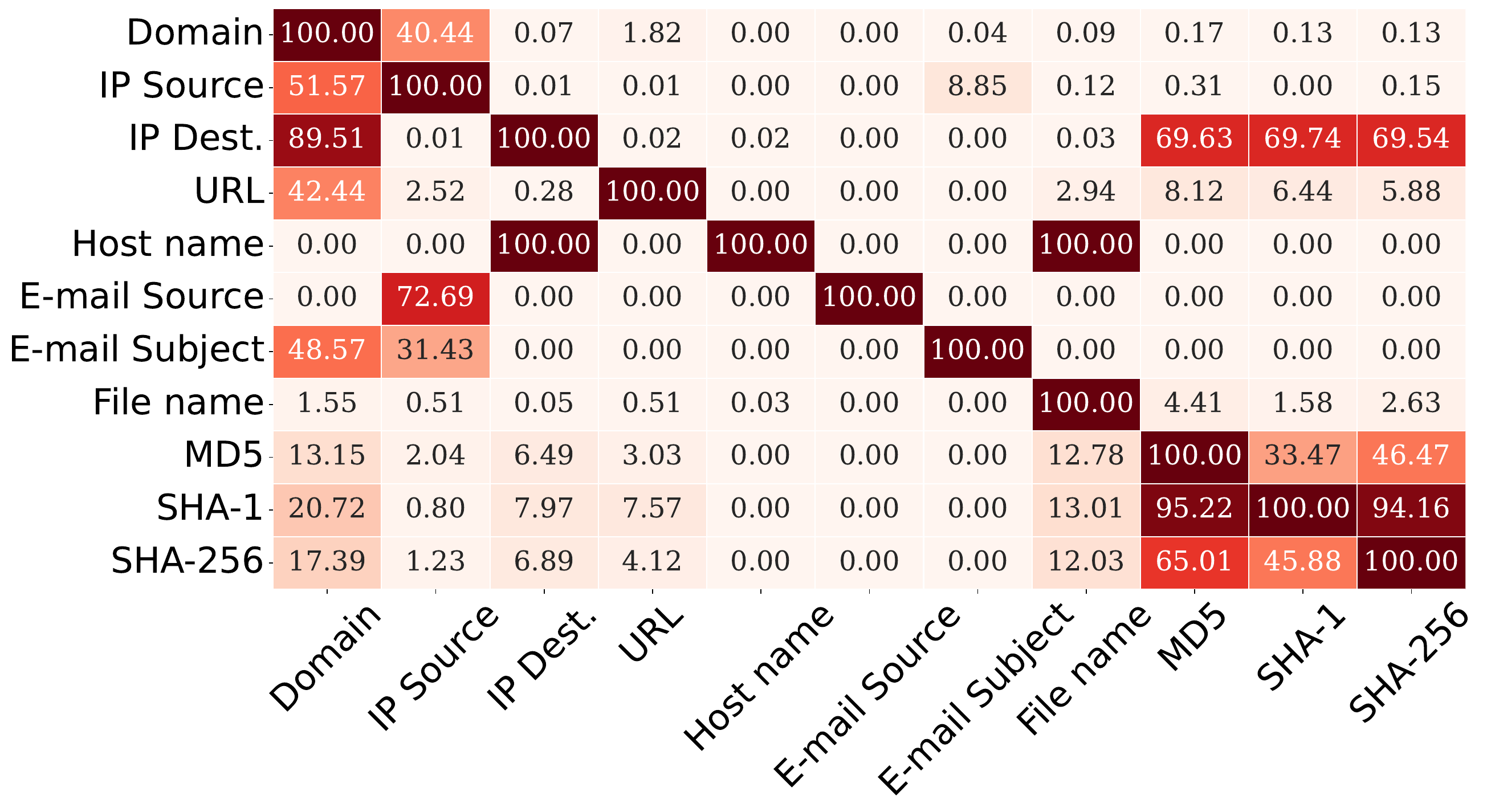} 
    \vspace{-0.15in}
    \caption{Correlation matrix of IOC categories, obtained from the relation between  IOC   sightings and events.  Each cell corresponds to the percentage of sightings pertaining to the row's category that appear in events that also have IOCs from the column's category.}
    \label{fig:correlation_matrix}
\end{figure}

Recall that  each sighting to an IOC belongs to an event, where an event  consists of a collection of sightings to related IOCs over time.  Despite the fact that the IOCs and events in our trace were anonymized, we are still able to correlate sighting categories through the corresponding events. To that aim, in Figure~\ref{fig:correlation_matrix}  each cell corresponds to the percentage of sightings pertaining to the row's category that appear in events that also present IOCs from the column's category. From this heat map, we can highlight two clusters.

The first cluster, in the bottom right corner, comprises MD5, SHA-1 and SHA-256 categories, in which we observe that whenever we have an indicator's MD5 hash, it is accompanied 33\% of the time by its SHA-1 hash and,  46\% of the time, by its SHA-256 hash. Indeed, it is typical to share different hash values for a given malware or  for its variants. The second cluster, in the top  left, suggests that domain names and IP addresses also tend to be shared through common events.

\begin{table}[t]
    \centering
    \caption{General statistics.} \label{tab:general} 
    \scalebox{1.0}{
    \begin{tabular}{l|r}
Metric & Value \\
   \hline
\hline
\# IOCs with at least one sighting &   {5,789  }\\
\# IOCs (total) &  $\approx 14,000,000$   \\
\#   sightings &  892,240\\
\hline
\#   IOCs with first sighting & \multicolumn{1}{c}{}   \\ before its creation date &  530    \\
\hline
 trace duration &  {724 days} \\
first creation date &  {2018-09-09} \\
last creation date &  {2020-09-02} \\
first sighting &  {2019-04-24} \\
last sighting &  {2020-09-02} \\  
\hline \hline
\multicolumn{2}{c}{Conditional metrics for IOCs with at least one sighting} \\
\hline
mean time to upcoming sighting, & \\
per IOC (multiple sightings & \\
can occur at same day) &  0.6 days \\
\hline
mean time to  first sighting, per IOC & \\
(after creation) & 83 days \\
average number of sightings, & \\
 per IOC & 154.13 \\
average number of days & \\
with sightings, per IOC & 2.72 days \\
\hline
\end{tabular}}
\end{table}

 \begin{figure}[b]
    \centering
    \includegraphics[width=0.6\columnwidth]{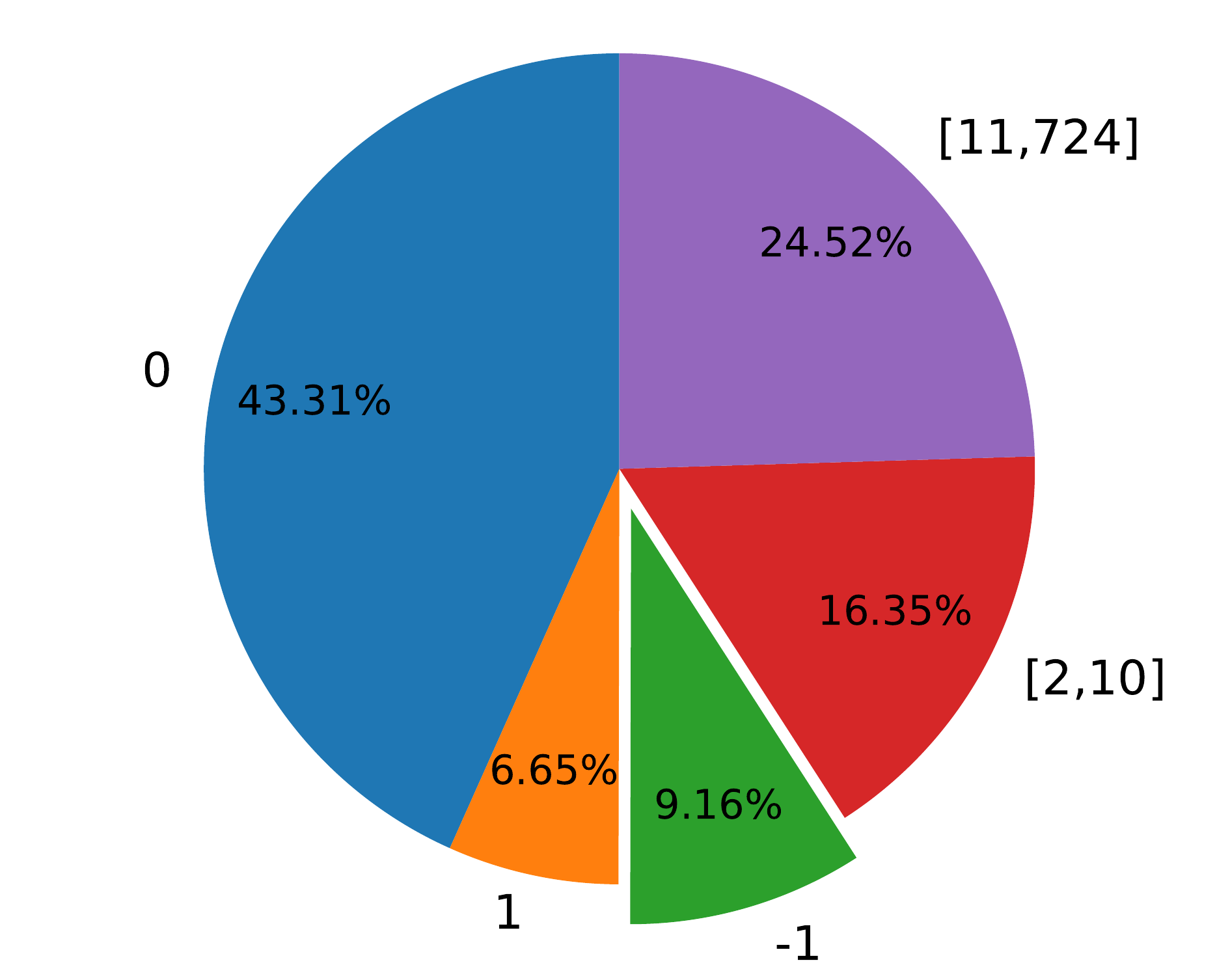}
    \caption{Distribution of first sighting minus creation date. Around 9\% of the IOCs have their first sighting one day before creation date, and   43\%   at the same day.}
    \label{fig:pie_chart}
\end{figure} 

\begin{finding}Hashes are typically sighted in bundles, and the same occurs for IPs and domain names.   The decaying models considered in this work can be used either to capture the aging of isolated IOCs or bundles of IOCs.  
\end{finding}


\section{Deterministic bounds and outliers} \label{sec:deterministicandoutliers}

\begin{researchquestion}
What is a  naive upper bound on TTL?
\end{researchquestion}

We begin by estimating an upper bound on the TTL value to cover all sightings for all IOCs. Indeed, a conservative approach towards IOC monitoring consists in setting TTL to a large enough value that, in retrospect, would have covered all sightings.  


In our trace, the largest gap between the first and last sightings towards an IOC equals 724 days.  Following a conservative monitoring strategy, i.e.,   $\timetolive=724$, and assuming an extension towards an infinite trace wherein  IOCs are created at a rate of $\lambda$ IOCs per time unit, it follows from Little's law the expected number of IOCs to be monitored at any point in time, in steady state, equals $724 \cdot \lambda$.   This estimate, however, is   sensitive to outliers, 
motivating the use of statistical tools to parameterize TTL in order to determine when and if IOCs should be  evicted. 


\begin{finding}
Setting TTL to the largest gap between IOC creation date and IOC sighting serves as a conservative upper bound on the TTL value, but this bound is   sensitive to outliers. \end{finding}

Whereas the above discussion accounted for a deterministic upper bound on   the TTL, in what follows  we consider a statistical perspective to account for outliers. 

\begin{researchquestion}
How to cope with outliers while trading between  monitoring costs and misses?
\end{researchquestion}

To cope with outliers and with the need for allowing a certain level of missed sightings, we consider statistical approaches to parametrize TTLs.  In the simplest setting, we take as inputs the target hit ratio $t$ (with corresponding miss ratio $1-t$) and the cumulative distribution function (CDF) of the time between consecutive sightings, $F(x)=P(X < x)$, where $X$ is a sample from the distribution of the time between sightings.  Then, we let $\timetolive=F^{-1}(t)$.

For large values of $t$, this model clearly degenerates to the simple deterministic bound discussed in the previous paragraph.  Smaller values of $t$ allow us to trade-off between coverage and monitoring costs. In our trace, to capture 90\% of sightings towards IOCs related to emails, we must let $\timetolive=38$  days.  In this case, a 10\% reduction in coverage corresponds to a 95\% decrease  ($1-38/724=0.95$)   in monitoring costs.

\begin{finding}
By reducing TTL, a small decrease in  coverage can yield a significant reduction in monitoring costs. This occurs in part due to outliers.
\end{finding}

\section{Accounting for categories} \label{sec:categories} 

\begin{researchquestion}
What is the impact of categories on IOC lifetimes?
\end{researchquestion}


In our trace we count with eleven IOC types: md5, sha1, sha256, ip-src, ip-dst, email-subject, email-dst, domain, hostname, filename and url. Each IOC is associated to exactly one type.  Conditioning TTL values to IOC types allows us to   reduce the impact of outliers, which skew the TTL values for the whole trace but may not impact certain categories.

\begin{figure}
    \centering
    \subfloat[Time between sightings.   
    Internal plot: zoom accounting for time between sightings of  up to 30 days. ]{
        \includegraphics[width=0.85\columnwidth]{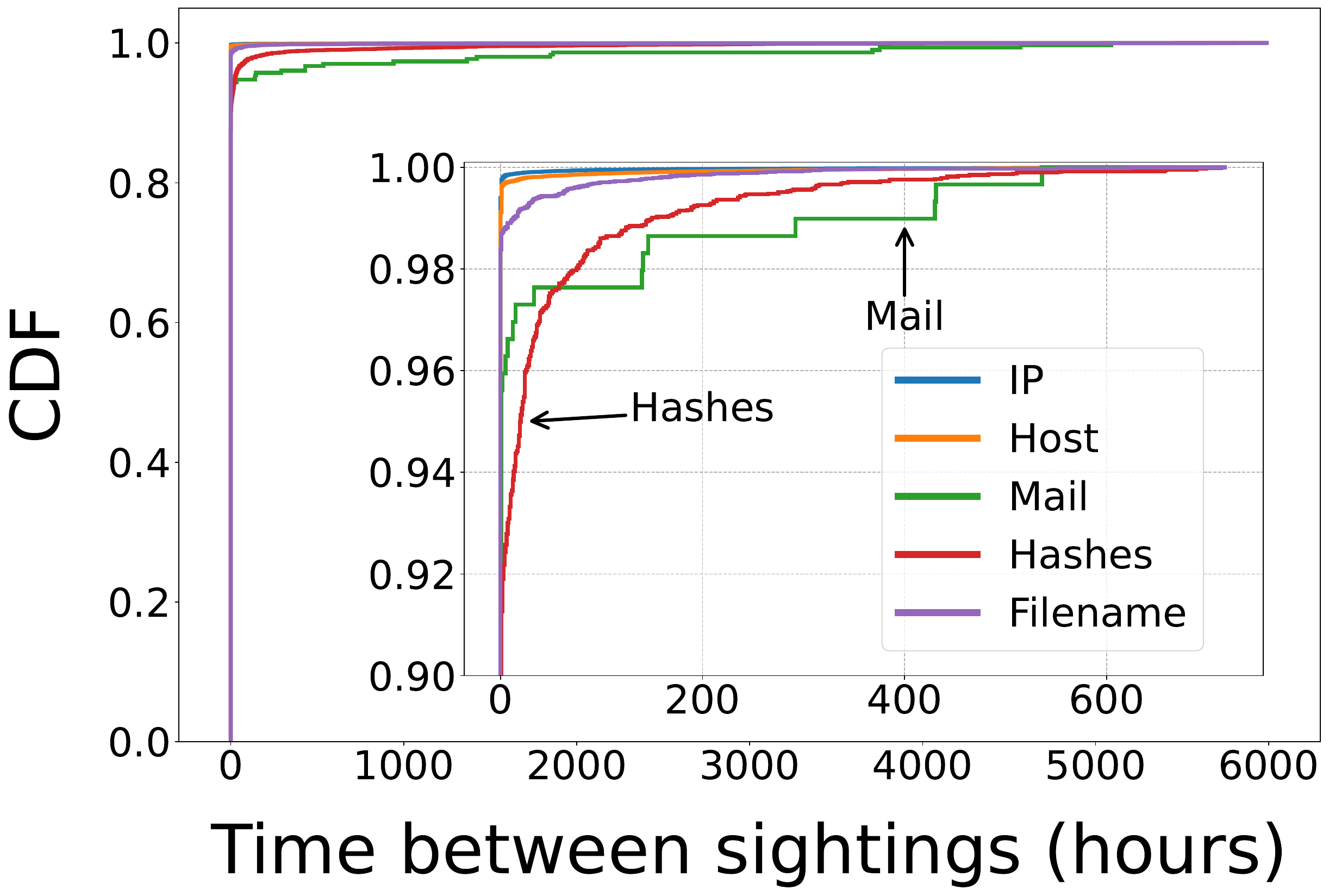}
        \label{fig:cdf_categorical_grouping1}
    }
    \hfill
    \subfloat[Time between  IOC creation date and  first sighting, measured in hours.]{
        \includegraphics[width=0.85\columnwidth]{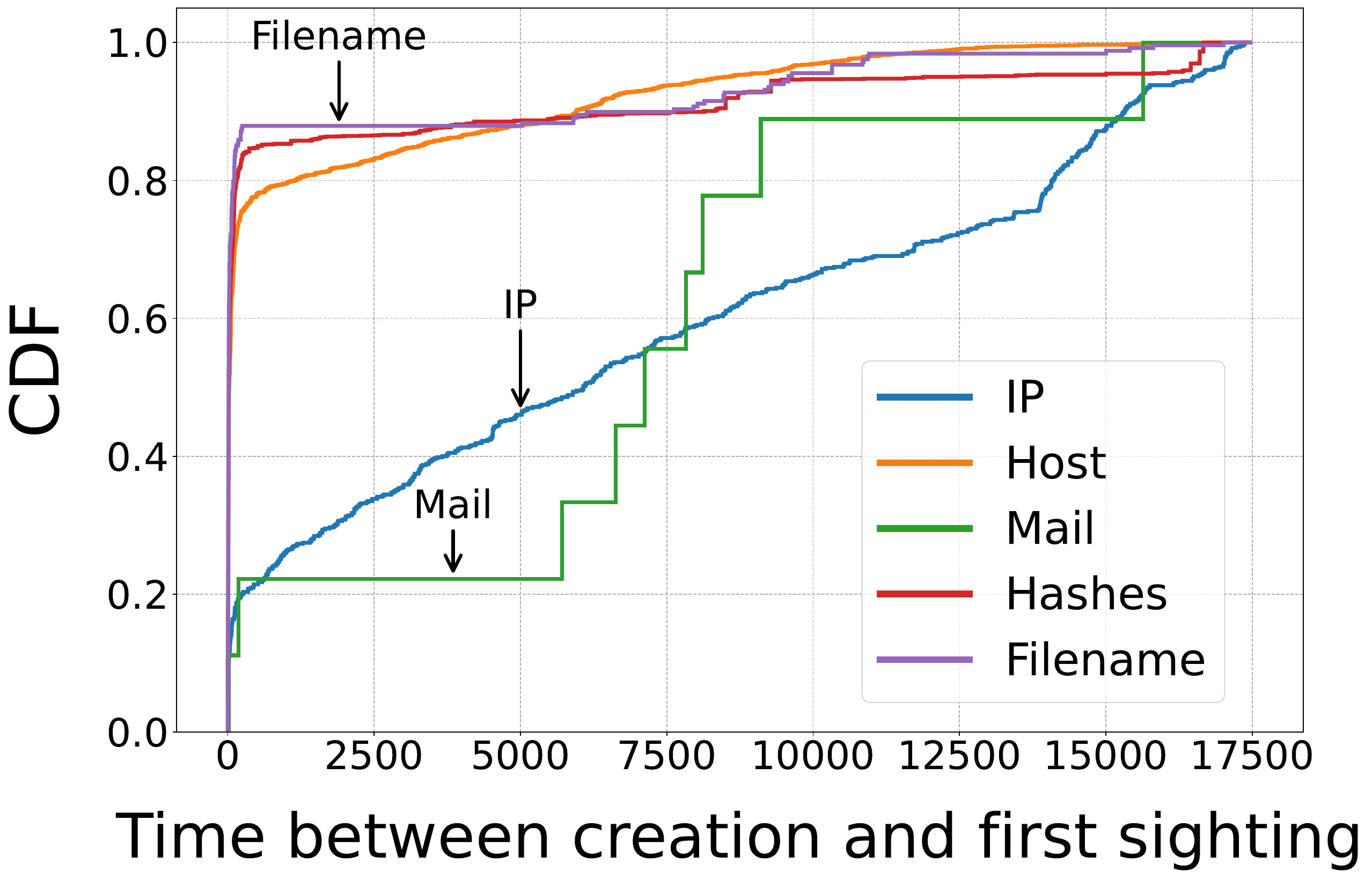}
        \label{fig:cdf_creation_to_fs}
    }
    \caption{Time between sightings and time until first sighting (measured in hours), over different IOC categories.  Hashes tend to linger   longer than IPs, showing larger times between sightings and times to first sightings.}
    \label{fig:acc_for_categories}
\end{figure}

The categories discussed above can be split or grouped.  As an example, the eleven categories may be grouped into five coarser clusters (see Figure \ref{fig:acc_for_categories}): hashes (md5, sha1, sha256), IPs (ip-src, ip-dst), email (email-subject, email-dst), host (domain, hostname, url) and filename.  In what follows, we refer to those five clusters of categories simply as \emph{categories}. 

   Figure \ref{fig:cdf_categorical_grouping1} 
   shows the cumulative distribution function (CDF) of the time between sightings, 
   for the five considered categories, respectively.  Hashes tend to linger longer than IPs, corresponding to  larger times between sightings. Indeed, while IPs are dynamic and should eventually be white-listed, hashes tend to be more stable over time.

Figure \ref{fig:cdf_creation_to_fs} shows the CDF of time from IOC  creation until first sighting, per category. 
For most categories, roughly 70\% of the IOCs have their first sighting soon after the creation date.  However, the remaining 30\% of IOCs have their first sightings uniformly distributed     throughout the period of observation. This means that for a significant fraction of IOCs the first sighting can take more than one year to occur.

\begin{finding}
There are at least two motivations to reduce TTL values: 1) reduce  monitoring costs and 2)   cope with IOCs that expire. The time between sightings   for hashes tends to be larger than for IPs.  This motivates setting TTLs for IPs to smaller values when compared against hashes, not only to reduce monitoring costs but also to cope with the fact that IP entries in blacklists  expire due to their ephemeral nature.  
\end{finding}




 \section{Utilities and costs} 

\label{sec:utilities}

\begin{researchquestion}
How to quantitatively capture the monitoring and miss costs while setting TTL?
\end{researchquestion}

 Next, we consider the availability of information about monitoring costs and costs associated with missing a sighting,  to determine the target hit ratio $t$.   Together, these costs can assist in the flexible monitoring of IOCs.  Indeed, such costs, e.g., measured in dollars per time unit and dollars per missed sighting, respectively, can be used to establish a utility function   to set  TTL  values in a principled utility-oriented  fashion.
 
 
Introducing monetary costs may impose additional challenges, as determining such costs is non-trivial.  However, monetary costs may help to convey the role of the IOC aging model in the considered organization, bridging the gap between IOC monitoring strategies and other elements of the business workflow.  Monetary costs can be determined exogenously based on related literature or on information provided by certain products, such as Azure Sentinel Threat Intelligence (ASTI). 

The ratio between monitoring and missing costs can also be estimated in an endogenous fashion, using data from the collected traces. Indeed, the trace of sightings implies that if the cost ratio is above a certain threshold, one should never monitor any IOC (no-monitoring extreme). At the other extreme of the spectrum, when the ratio is below a lower threshold all IOCs should be constantly monitored  (always-monitoring extreme).  Knowing such two thresholds, and understanding how the cost ratio impacts monitoring strategies, together with historical information about monitoring practices in a given business, provides insights on the current and prospective target cost ratios.

To find the two thresholds referred to in the above paragraph, we define TTLs ranging between 0 and the maximum interval between sightings (see Section~\ref{sec:deterministicandoutliers}). For each TTL value we compute, in retrospect,  the corresponding monitoring and missing costs.  The monitoring cost is the number of days we monitor each IOC in our system multiplied by the cost of each day of monitoring.   The missing cost is the number of missed sightings multiplied by the cost of each miss.

Let $C$ be the total cost, and $\costmon$ and $\costmiss$ be the monitoring and missed sighting costs, respectively.  $\costmon$  is the monitoring cost per IOC per day, and $\costmiss$  is the miss cost per missed sighting.   Under the above simple model, the total cost is a linear function of the time that IOCs were monitored (accounting for  IOCs with no sighting) and the number of missed sightings. Let $I$ be the number of IOCs:
\begin{align}
&C(\nmon, \nmiss; \costmon, \costmiss)=\\
& \quad=\costmon \sum_{i=1}^I \nmoni  + \costmiss \sum_{i=1}^I \nmissi \\
&\quad = \costmon \nmon  + \costmiss \nmiss
\end{align}
where $\nmoni$ and $\nmissi$ are the monitoring time and number of missed sightings for the $i$-th IOC, and $\nmon$ and $\nmiss$ are the corresponding quantities accounting for all IOCs.  Note that $\nmon$ and $\nmiss$ are functions of $\timetolive$. Indeed, as many IOCs receive no sightings, we have
\begin{align}
    &N_{\textrm{mon}}^{(i)}= \nonumber \\
    &=\left\{ \begin{array}{ll}
        T, & \textrm{if IOC } i \textrm{ has no sightings} \\
        T+\Delta T^{(i)}, & \textrm{otherwise} 
    \end{array} \right.
\end{align}
where $\Delta T^{(i)}$ is the additional number of days at which IOC $i$ is monitored beyond $T$.  Then,
\begin{equation}
N_{\textrm{mon}} = T \cdot I + \sum_{i=1}^I    \Delta T^{(i)} \label{eq:costmon}
\end{equation}
where $\Delta T^{(i)}=0$ for TTL without reset, and can be positive for TTL with reset.

The dependency of $\nmon$ and $\nmiss$ on $T$ may be non-trivial, e.g., non-convex.  In what follows,  we denote the functions that  capture such dependencies and that map $T$ into $\nmon$ and $\nmiss$ as $g(\cdot)$ and $h(\cdot)$,
\begin{align}
    \nmon &= g(T) \\
\nmiss &= h(T)
\end{align}
We first provide some back-of-the-envelope calculations to get some insights on how cost impacts the optimal TTL, and the 
proceed with a  trace-driven exploration to 1) determine the best TTL value, given the costs of monitoring and missing and 2) search for the two cost ratios that correspond to the extremal thresholds discussed above.   

\begin{finding}
The impact of monitoring and miss costs can be captured through an utility function consisting of the sum of   two costs, which vary in a non-linear fashion with respect to TTL values. To deal with  such non-linearity,   one alternative is to approach the problem of finding the optimal TTL through a trace-driven perspective. 
\end{finding}

  \subsection{Back-of-the-envelope calculations}
  
  \begin{researchquestion}
  Is it feasible to get  rough estimates of recommended TTL values without leveraging the full trace of sightings in its details?
  \end{researchquestion}
  
  Next, we aim at providing initial insights on the proposed cost model, leveraging data from our traces.
  We begin by revisiting Equation~\eqref{eq:costmon}.  In particular, we consider a simple  workload model wherein  sightings  to IOC $i$  arrive according to a Poisson process with rate $\lambda_i$.  Then, under  TTL with reset, the probability that the TTL is not reset during an interval of $T$ seconds is given by $\exp(-\lambda_i T)$, which corresponds to  the probability that no sighting arrives during that interval. In that case, assuming that every reset increases the monitoring period by $T/2$, we have
  \begin{equation}
      \Delta T^{(i)} = T e^{\lambda_i T}/2.
  \end{equation}
According to the above model, the cost of monitoring increases exponentially with $T$. However, our trace-driven analysis shows that in practice, the monitoring cost increases linearly as $T$ increases. This is because we observed that sightings of a particular IOC tend to occur in bursts, meaning that many sightings occur on the same day. To estimate the arrival rate of sightings that contribute to resets, we count all sightings in a burst as a single burst arrival. 
We found that the average number of daily bursts of sightings per IOC is 2.72 (as shown in the last line of Table~\ref{tab:general}),  which is quite small compared to the overall trace duration. As a result, $\lambda_i$ approaches zero, causing the monitoring cost to increase roughly linearly with respect to $T$ in the scenarios   examined in the following section.

Next, we provide   rough estimates of two thresholds on the ratio between monitoring costs and missed sighting costs,  according to which it is beneficial never to monitor any IOC or always to monitor all IOCs.  Let \begin{equation}
    R=\frac{\costmon}{\costmiss}. \label{def:r}
\end{equation} For the first threshold, that we refer to as $R_U$, we have that
\begin{align}
    R \ge R_U \Rightarrow T^{\star} = 0
\end{align}
where $T^{\star}$ is the optimal threshold.  Similarly, for the second threshold, referred to as $R_L,$
\begin{align}
    R \le R_L \Rightarrow T^{\star} = \tilde{T}
\end{align}
where $\tilde{T}$ is the maximum admitted threshold. 

In the following section, we formally pose optimization problems that yield the above thresholds, and proceed with a trace-driven simulation to determine $R_L$ and $R_U$. The results are shown in Table~\ref{tab:basic_statistics1}. Alternatively, in what follows we provide some back-of-the-envelope  heuristics  to approximate those thresholds.







 The total number of sightings in our trace is $892,240$, spread over 724 days and across roughly 14 million IOCs, most of which having no sightings. 
 Assuming a normalized unitary monitoring cost per IOC per day, the monitoring cost to monitor a single IOC during the interval  of interest will be of 724 normalized monetary units.  The gain, in contrast, will be on average of $\costmiss \cdot 892,240 / (14 \cdot 10^6)$, assuming that the probability that the monitored IOC will be sighted is proportional to the fraction of IOCs that receive at least one sighting. In particular, in our trace we have an average of $892,240 / 5,789$  sightings per IOC, conditioned on IOCs that have at least one sighting,  and a fraction of $5,789/(14 \cdot 10^6)$ IOCs that are sighted.  Multiplying the two quantities, we obtain the expected number of sightings covered by an IOC.
 Therefore, it is worth monitoring at least one IOC per  day
 if  $\costmon \cdot 724 < \costmiss \cdot 892,240/(14 \cdot 10^6)$. 
 Letting $\costmon=1$, if $ 1/\costmiss > 892,240/(14 \cdot 10^6 \cdot 724) = 1/11,363$ it is not worth monitoring even one IOC per day. As shown in Table~\ref{tab:basic_statistics1}, indeed such ballpark value is on the same order of magnitude of 1/2,115 that we found in our trace-driven evaluations as the  cost ratio   above which   monitoring is not worthy.
 
On the other extreme of the spectrum, if the cost of a miss at a given day, $\costmiss$, is larger than the cost of monitoring all IOCs during that day, one should always monitor all IOCs.  Given that we have roughly $14$ million IOCs, this amounts to always monitoring if  $\costmiss > 14 \cdot 10^6$.  Again, the order of magnitude is in agreement with the results presented in Table~\ref{tab:basic_statistics1}. The difference between the values in Table~\ref{tab:basic_statistics1} and the above back-of-the-envelope calculations are due to a number of factors, including 1) the fact that  IOCs are created in a non-uniform fashion over the trace and  2)  sightings tend to occur in bursts. To capture those details, we perform a trace-driven evaluation as further discussed in the next section.

\begin{finding}
The ratio between monitoring costs and miss costs plays a key role in determining when it is optimal to never monitor any IOC, or to always monitor all IOCs.   Those two strategies are optimal when the miss cost is on the order of thousands and millions, respectively, when compared against   unitary monitoring costs.
\end{finding}

 

\subsection{Model and optimization problem}

\begin{researchquestion}How does a detailed trace analysis compare against back of the envelope calculations to determine optimal TTL values?
\end{researchquestion}

The optimization problem corresponding to the optimal TTL estimation is given as follows:
 \begin{align}
    \timetolive^*: \quad  &		\textrm{Argmin}_{\timetolive} & 	C(\nmon, \nmiss; \costmon, \costmiss) \nonumber \\  
& \textrm{Subject to} \nonumber   &	\nmon=g(\timetolive) \\
&		&		\nmiss=h(\timetolive)  \nonumber 
\end{align}

Recall that $R=\costmon/ \costmiss$ (see Equation~\ref{def:r}). Then, \begin{align}
    &\textrm{Argmin}_{\timetolive} 	C(\nmon, \nmiss; \costmon, \costmiss) = \nonumber \\
    &\quad =\textrm{Argmin}_{\timetolive} R\cdot g(T) + h(T) \label{eq:newargmin}
\end{align} 
In Appendix~\ref{sec:appa} we specialize the above optimization problem to determine $R_U$ and $R_L$, i.e., the thresholds beyond which IOCs should never be monitored, or should be always monitored, respectively.

Note that in the above formulation we assumed that functions $g(\cdot)$ and $h(\cdot)$ are obtained  from traces.  Alternatively, approximating those functions through simple expressions may be instrumental to express the solutions to the above problems in closed-form, which we leave as subject for future work.

\begin{table}[t]
    \centering
    \caption{Trace-driven bounds. The lower bound $R_L$ corresponds to the ratio monitoring cost:miss cost  motivating always monitoring, i.e.,   setting $T=724$  days. The upper bound $R_U$ corresponds to the ratio    motivating never monitoring, i.e., setting $T=0$.} \scalebox{1.0}{
    \begin{tabular}{c|r|r}
\hline \hline
& \multicolumn{2}{c}{TTL} \\
& {with reset} & {without reset} \\
        \hline
$R_L$     &  1:10,504,881 &  1:843,470   \\
\hline
$R_U$   &   1:2,152  &   1:2,115   \\
\hline
    \end{tabular} }
    \label{tab:basic_statistics1}
\end{table}

\begin{figure}
    \centering
    \includegraphics[width=0.85\columnwidth]{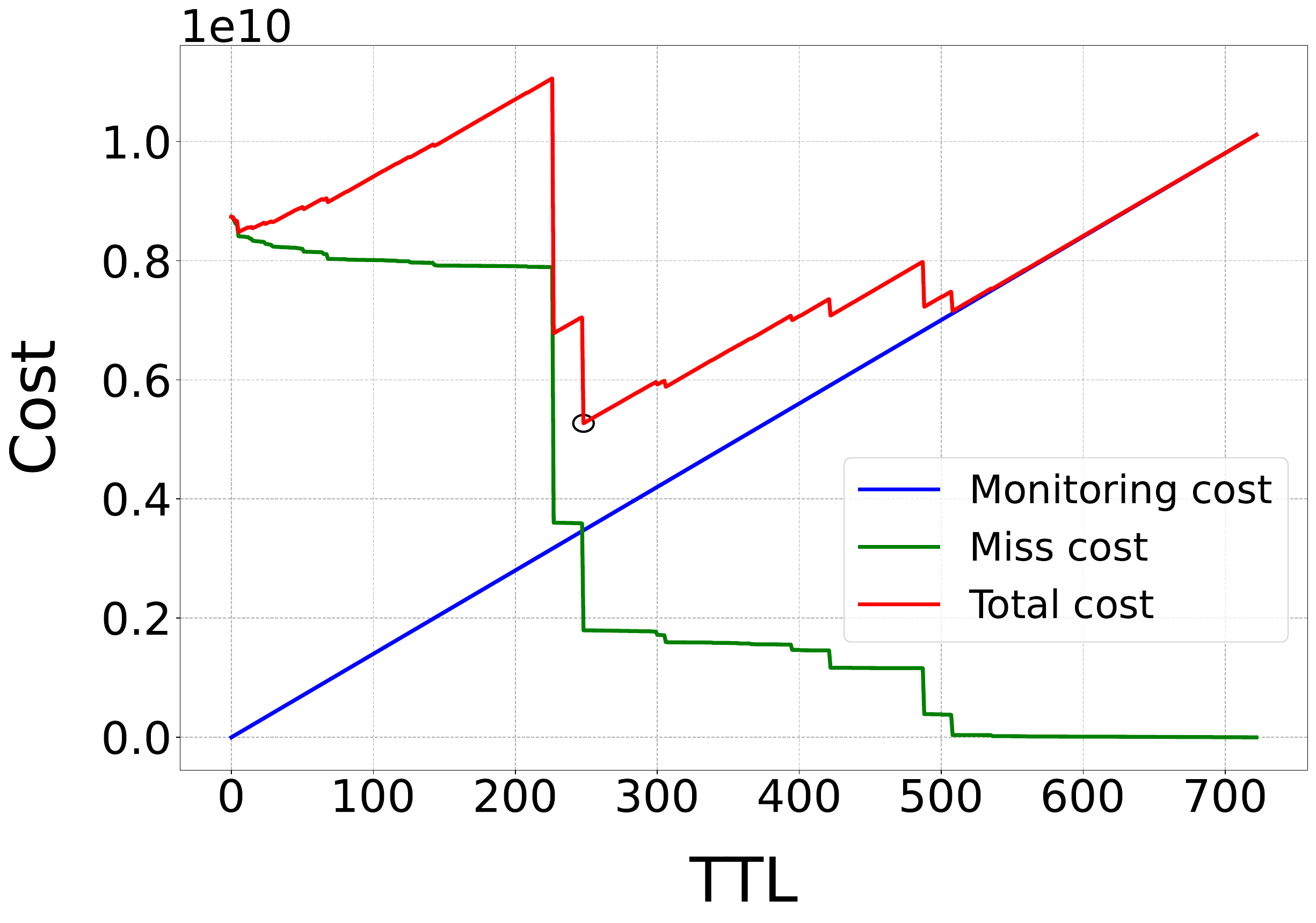}
    \caption{Costs as a function of $T$ for $R=\costmon/\costmiss = 1/10,000$. The monitoring cost $\costmon \cdot \nmon$ grows with respect to $T$, whereas the miss cost $\costmiss \cdot \nmiss$ decreases. The sum of the two costs, $C$, has its minimum  at $T=248$.}
    \label{fig:cost_curves}
\end{figure}

Table~\ref{tab:basic_statistics1} shows the values of $R_L$ and $R_U$ obtained as solutions of the above optimization problems. Leveraging our traces, we learn that if the  cost of a miss is on the order of millions when compared to the   monitoring cost per IOC per day, it is worth always monitoring all IOCs. Alternatively, if the miss cost is around 2,000 times the  monitoring cost per IOC per day, it is not worth monitoring any IOC.  For intermediary cost values, Figure~\ref{fig:cost_curves} shows how the cost varies as a function of TTL.


Figure \ref{fig:cost_curves} shows  how the monitoring cost, $\costmon \nmon$, miss cost, $\costmiss \nmiss$, and total cost, $C$, varies as a function of TTL, $T$.  
We let $\costmon = 1$ USD and $\costmiss = 10,000$ USD, i.e.,  $R = {1}/{10,000}$. The curve is obtained using TTL with reset.   Note that  the monitoring cost grows roughly linearly with respect to $T$, as previously discussed under our back-of-the-envelope calculations. In addition, note that the miss cost decreases with respect to $T$, showing a steep decrease around $T=200$.  Finally, the optimal TTL equals 248, and occurs around the point where the cost curves corresponding to monitoring and miss costs cross each other.   


\begin{figure}
    \centering
    \includegraphics[width=0.85\columnwidth]{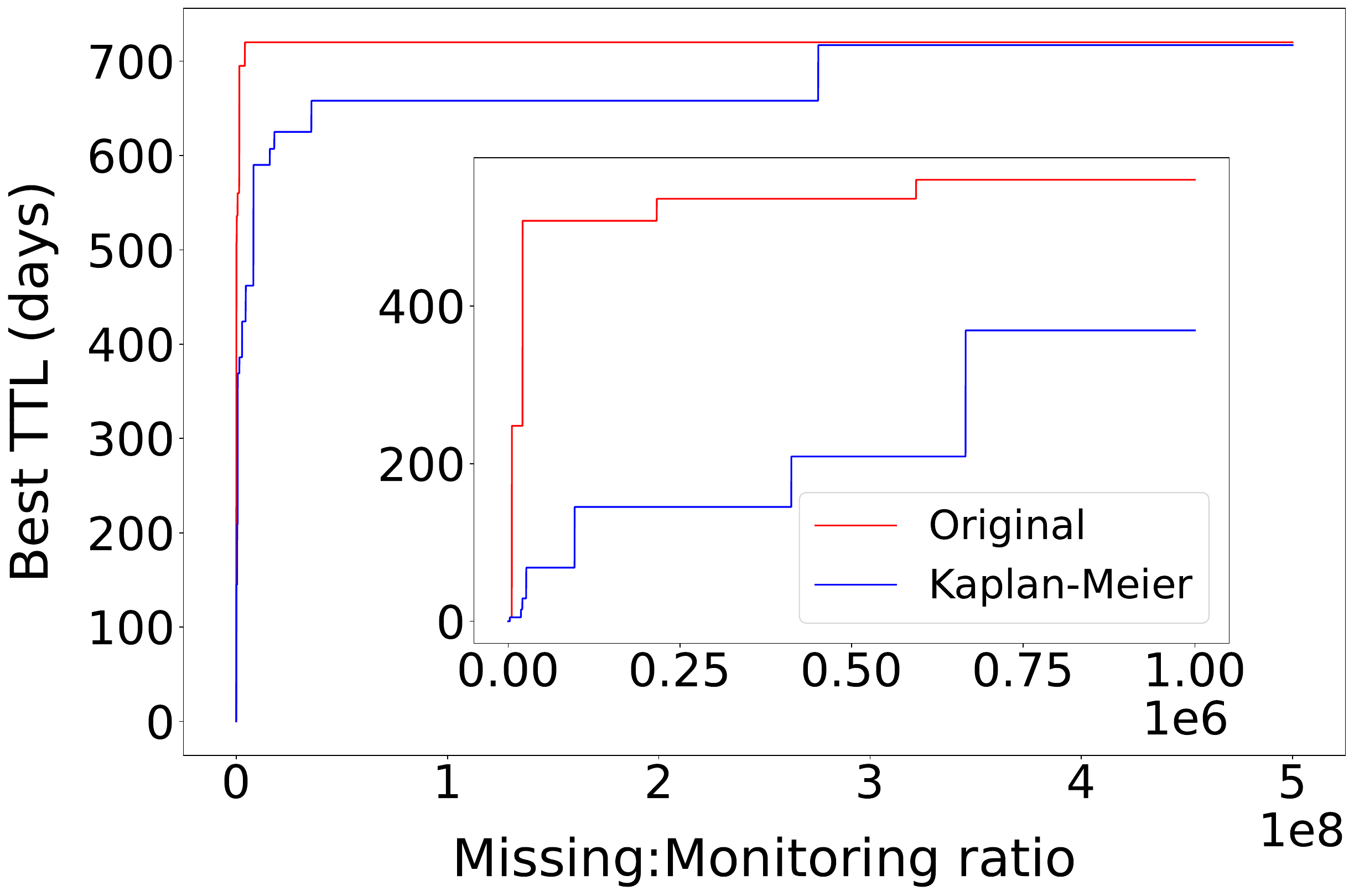}
    \caption{Best TTL as a function of        miss   over  monitoring cost ratio ($R^{-1}=\costmiss/\costmon$), under TTL with reset. }
    \label{fig:best_TTLs}
\end{figure}

We repeat the above methodology for $\costmiss$ varying between 1 and $10^8$, keeping $\costmon=1$.  The results are reported in Figure~\ref{fig:best_TTLs}, that shows how the best $T$ varies as a function of the ratio $R^{-1}=\costmiss/\costmon$. The figure shows that as $\costmiss$ increases from 10,000 to 100,000, the optimal TTL value rapidly grows from 248 (as discussed in the previous paragraph) to 500 (as shown in the inside zoom box in Figure~\ref{fig:best_TTLs}).  The maximum TTL value is     724 days,  which corresponds to the trace duration, and is reached when the miss cost reaches roughly 10 million times the monitoring costs (as previously reported in Table~\ref{tab:basic_statistics1}).

\begin{finding}
The trace-driven analysis is in agreement with the back-of-the-envelope calculations, and further allows us to assess the optimal TTL values in the range between 0 and 724 days when  the miss costs vary from the order of thousands to the order of million times the normalized monitoring costs.
\end{finding}

\section{Conclusion} \label{sec:conclusion}

Platforms for information sharing are key elements of the cyber-security ecosystem~\cite{miranda2021flow,preuveneers2021sharing,sridhar2021cybersecurity}. By facilitating the exchange of IOCs and sightings among stakeholders, such platforms allow SOC employees to prioritize which IOCs must be monitored and which alerts require immediate attention.  Despite the tremendous success of platforms of that sort, such as MISP, and the proposed features  to capture the aging of IOCs, their users are still challenged by the parametrization of the aging  models.  In particular, such platforms allow users to share sightings, but sensitive information about how those sightings impact monitoring is typically kept private. 

In this work, we provide initial insights obtained from a flow of sightings towards IOCs in a real world environment. 
By leveraging a trace of sightings towards IOCs, we  consider the problem of determining optimal schedules for the monitoring of  IOCs. 
While considering the costs of monitoring IOCs and of missed sightings, we  provide a methodology to determine lower and upper bounds ($R_L$ and $R_U$, respectively) on the ratio between such costs so that IOCs should be   monitored  forever or not monitored at all. Such ratio, together with traces collected from a SOC, allow operators to tune TTL values according to their needs.  We illustrate the methodology using a trace collected from a real world environment, and envision that the methodology is applicable across different businesses.

 This work opens up a number of avenues related to IOC decaying models. First, we simplified the parameterization of an IOC decaying model, mapping it into  the assessment of  the ratio between miss costs and monitoring costs, which we believe is  close to the reality of operators.   Further assessing how that ratio varies across environments is left as subject for future work.  Second, we provided a trace-driven solution to the problem of determining the best TTL.  Analytical solutions, e.g., leveraging special properties on how the number of misses or the monitoring costs vary as a function of the TTL, are also left as subject for future work.  
 
 \textbf{Acknowledgment: }
 This work was supported in part by SIEMENS,   CAPES,   CNPq, and  FAPERJ under Grants 315110/2020-1, E-26/211.144/2019, and E-26/201.376/2021. 
 
 


\bibliographystyle{IEEEtran}
\bibliography{main}

\appendices

\section{Workload Characterization}
\label{sec:appworkload}

We provide further insights on the workload of our traces.  

\subsection{Curve fitting}

\begin{researchquestion}
Does the time between sightings follow a power law?
\end{researchquestion}

We plotted the complementary cumulative distribution function (CCDF)  of time between all sightings in our trace, and observed a  linear decay       typical of a power law~\cite{Newman_2005}.   A similar power law behavior for time between   security incidents was observed in~\cite{frei2006largescale}.    Motivated by these observations, 
we selected Pareto and Weibull distributions to fit our data. 
The Pareto and Weibull CCDFs are given by
$ 
\overline{F}(x) =  \left({k}/{x}\right)^\alpha $  and
$
\overline{F}(x) =   e^{-(x/k)^{\lambda}}
 $,  respectively. 
The parameters of the best fits for each category are listed in Table~\ref{tab:pareto_fits},   
for the Pareto and Weibull fits. 
Table \ref{tab:pareto_fits} also provides   insight on the quality of     fits. For instance,     IOCs corresponding to hashes (MD5, SHA-1 and SHA-256) tend to be better captured through a Weibull distribution as opposed to other IOCs (domains and IPs)  that find a good fit   under both Pareto and Weibull.

\begin{table}[t]
    \centering
    \caption{ Pareto and Weibull fits to time between sightings.}
      \scalebox{0.75}{
\begin{tabular}{|c|c|c|c|c|}
\hline
\textbf{Category} & \textbf{Weibull Parameters}        & \textbf{  RMSE} & \textbf{Pareto Parameters}    & \textbf{  RMSE} \\ \hline \hline
Domain   & $\lambda = 0.2057, k = 0.0004$ & $1e^{-5}$ & $\alpha = 1.2, k = 0.0199$ & $8e^{-6}$ \\ \hline
IP Source         & $\lambda = 0.1191, k = 9.65e^{-7}$ & $1e^{-5}$         & $\alpha = 0.8, k = 0.0020$ & $2e^{-5}$        \\ \hline
IP Destination    & $\lambda = 0.1354, k = 3.01e^{-5}$ & $5e^{-5}$         & $\alpha = 0.7, k = 0.0067$ & $6e^{-5}$        \\ \hline
MD5      & $\lambda = 0.2731, k = 0.2983$ & $1e^{-3}$ & $\alpha = 0.7, k = 0.1914$ & $2e^{-3}$ \\ \hline
SHA1     & $\lambda = 0.2735, k = 1.3974$ & $3e^{-3}$ & $\alpha = 0.6, k = 0.3322$ & $7e^{-3}$ \\ \hline
SHA256   & $\lambda = 0.2990, k = 0.6838$ & $1e^{-3}$ & $\alpha = 0.7, k = 0.2623$ & $3e^{-3}$ \\ \hline
Filename & $\lambda = 0.1375, k = 0.0003$ & $5e^{-4}$ & $\alpha = 0.6, k = 0.0097$ & $7e^{-4}$ \\ \hline
URL      & $\lambda = 0.2978, k = 2.2679$ & $1e^{-5}$ & $\alpha = 0.6, k = 0.4199$ & $9e^{-3}$ \\ \hline
\end{tabular} }
\label{tab:pareto_fits} \vspace{-0.2in}
\end{table}

\begin{finding}
The time between sightings follows a power law, whose parameters are presented in Table~\ref{tab:pareto_fits}.
\end{finding}

\subsection{Survival analysis}

\begin{researchquestion}How to cope with the fact that many IOCs, for which the exact creation date is unknown,  have their creation dates set at the very beginning of our trace?
\end{researchquestion}

The creation dates of   IOCs in our trace are lower bounded by September 9, 2018. Around 10\% of the  IOCs (533 IOCs) were created before that date, but  are attributed such lower bound as a reference   (see Figure~\ref{fig:cdf_IOC}).  Under survival analysis, this sort of reference   attribution is referred to as censoring.   In what follows, we apply the standard  Kaplan-Meier (KM) survival estimator  to adjust the dataset, filling up missing data for creation dates that occurred before  the reference. 

\begin{figure}[t]
    \centering
    \includegraphics[width=0.7\columnwidth]{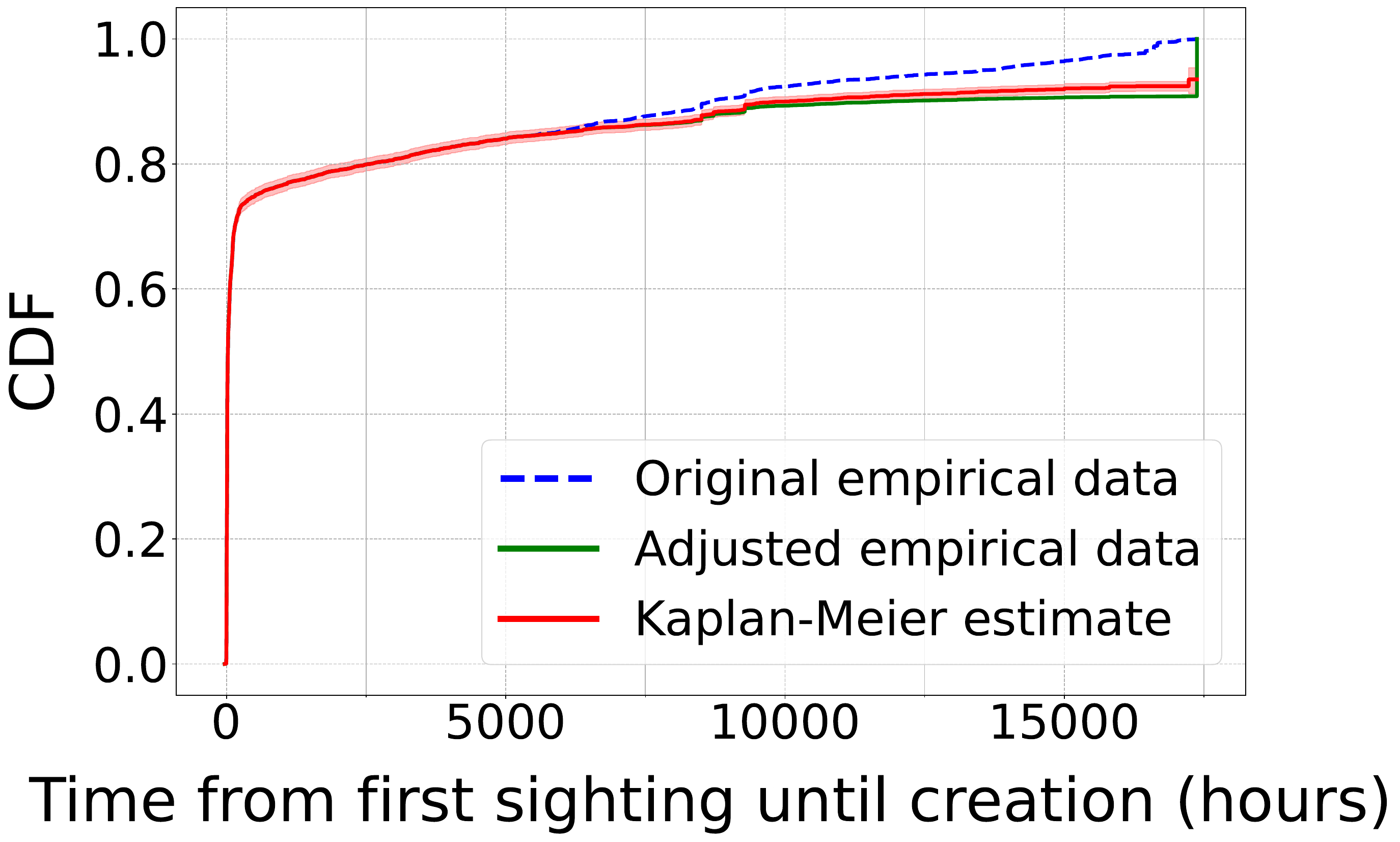}
    \caption{CDF of time between creation   and first sighting. In  dotted blue, the empirical CDF;  in red,  the Kaplan-Meier (KM) estimator; in blue, a  simple adjustment to the empirical values  that produces results similar to KM.}
    \label{fig:kme_cd_to_fs}  
\end{figure}

Figure~\ref{fig:kme_cd_to_fs} shows the  results obtained through the KM estimator.   The dotted blue line shows the empirical CDF of time to first sighting, and the red line shows the adjusted CDF obtained using the KM estimator.  The green line corresponds to a simple heuristic, wherein for each     IOC $i$ for which the creation date equals  the reference value of September 9, 2018, we replace its creation date by its first sighting minus 724 days, which corresponds to our trace duration. The rationale is that, for the oldest IOCs,  the trace duration is a good proxy for the time until  first sighting. Indeed, Figure~\ref{fig:kme_cd_to_fs} indicates that this heuristic closely captures the behavior of KM, and we adopt it to reassess  optimal TTL values in light of survival analysis, referring to it simply as KM.

Figure \ref{fig:best_TTLs} indicates how the   adjustment of  creation dates impacts optimal TTLs.  Note that after   distancing  creation dates from the IOC first sighting, additional time is spent monitoring IOCs while   no sightings are observed.  This, in turn,  makes monitoring less rewarding, favoring smaller TTL values.

\begin{finding}
Survival analysis provides a methodology to cope with missing data, and alllows us to revisit the optimal TTL values in light of missing data. In particular, after filling up missing data, optimal TTL values   increase, given that   unknown creation dates  occurred   before the beginning of the trace.
\end{finding}

\section{Details on optimization problem} \label{sec:appa}

First, we consider the problem  of determining an upper bound on the cost ratio $R_U$ beyond which IOCs should never be monitored. Given~\eqref{eq:newargmin}, the optimization problem is:
 \begin{align}
R_U: &\textrm{Min} \quad 	R \nonumber \\
&  \nonumber 
 \textrm{Subject to } \\
 &\quad \textrm{Argmin}_{\timetolive} R  \cdot g(\timetolive)+h(\timetolive) =0  \nonumber  
\end{align}
Note that $g(0)=0$, therefore for $R > R_U$ we have $C= \costmiss \cdot \nmiss = \costmiss   \cdot h(T)$. In the regime wherein monitoring costs are high, no IOCs are monitored and the ultimate cost depends only on the number of missed sightings.

Correspondingly, the optimization problem to   determine a lower bound on the cost ratio $R_L$ below which IOCs should always be monitored is:
\begin{align}
R_L: &\textrm{Max} \quad 	R \nonumber \\
&  \nonumber 
 \textrm{Subject to } \\
 &\quad  \textrm{Argmin}_{\timetolive} R  \cdot g(\timetolive)+h(\timetolive) = \widetilde{ \timetolive}  \nonumber  
\end{align}
%
%
where    $\widetilde{\timetolive}$   is the maximum feasible TTL value.  
Under a trace-driven approach,   $\widetilde{\timetolive}$   can be set  as the maximum interval between sightings (see Section~\ref{sec:deterministicandoutliers}). 
Assuming $h(\widetilde{\timetolive})=0$, i.e., no sightings are missed when $\timetolive{}$ is set to $\widetilde{\timetolive}$, condition $R < R_L$ implies that $C/\costmiss=R \cdot  g(\timetolive)$. In the regime wherein monitoring costs are low, all IOCs are monitored and the ultimate cost depends only on the product $\costmon \cdot \nmon = \costmon \cdot  g(\timetolive)$.






\begin{IEEEbiography}
{Breno Tostes}  received his B.Sc. degree in Computer Science from Federal University of Rio de Janeiro (UFRJ) in 2022.  His interests include modeling and analysis of computer systems, with focus on performance evaluation and cybersecurity. 
\end{IEEEbiography}

\begin{IEEEbiography}
{Leonardo Ventura}  received his B.Sc. degree in Computer Science from Federal University of Rio de Janeiro (UFRJ) in 2022. He is currently a software engineer at Uber, with focus on cybersecurity. 
\end{IEEEbiography}

\begin{IEEEbiography}
{Enrico Lovat}  received his PhD from the Technical University of Munich for his research on the topics of usage control and information flow tracking. He joined Siemens CERT in 2016 as incident handler. Currently, he is responsible for Cyber Threat Intelligence at Siemens CERT.
\end{IEEEbiography}

\begin{IEEEbiography}
 {Matheus Martins} 
    received the B.S. and M.Sc. degrees in Computer Science and Informatics from Federal University of Rio de Janeiro (UFRJ), in 2018 and 2020, respectively.   His research focuses on data science to create cybersecurity risk models for patch management. Currently, he is working with cyber threat intelligence and incident response at Siemens Technology.  His interests include modeling and analysis of systems, with focus on cybersecurity.
\end{IEEEbiography}

\begin{IEEEbiography}
 {Daniel Sadoc Menasch\'e} 
    received the Ph.D. degree in Computer Science from the University of Massachusetts, Amherst, in 2011. Currently, he is an Associate Professor with the Computer Science Department, Federal University of Rio de Janeiro, Brazil. His interests are in modeling, analysis, security and performance evaluation of computer systems, being a recipient of best paper awards at GLOBECOM 2007, CoNEXT 2009, INFOCOM 2013 and ICGSE 2015.  He is an alumnus affiliated member of the Brazilian Academy of Sciences.
\end{IEEEbiography}

\end{document}